\documentclass{svproc}
\usepackage{url}
\usepackage{graphicx}

\begin{document}
\mainmatter              
\title{Recent Extensions of the ZKCM Library for Parallel and Accurate MPS Simulation of Quantum Circuits}
\titlerunning{Recent Extensions}  
%
\author{Akira SaiToh}
\authorrunning{Akira SaiToh} 
%
\tocauthor{Akira SaiToh}
\institute{Department of Computer and Information Sciences, Sojo University,\\
4-22-1 Ikeda, Nishi-ku, Kumamoto, Kumamoto 860-0082, Japan,\\
\email{st@cis.sojo-u.ac.jp}~\\~\\
{\rm Submitted to the post-conference proceedings of CCP2023 on 30 Sep. 2023.}}
\maketitle              

\begin{abstract}
A C++ library ZKCM and its extension library ZKCM\_QC have been developed since
2011 for multiple-precision matrix computation and accurate matrix-product-state
(MPS) quantum circuit simulation, respectively. In this report, a recent progress
in the extensions of these libraries is described, which are mainly for parallel
processing with the OpenMP and CUDA frameworks.
\keywords{parallel processing, multiple-precision computing, matrix product state, quantum circuits}
\end{abstract}
\section{Introduction}
A C++ library ZKCM \cite{AST-ZKCM} has been developed for multiple-precision matrix computation
since 2011. Its extension library ZKCM\_QC \cite{AST-S13-1,AST-S13-2,AST-S14} was developed for
a matrix-product-state (MPS) simulation \cite{V03} of quantum algorithms \cite{NC2000}.
The MPS simulation method is known as one of effective simulation techniques for classically simulating
quantum computing \cite{V03,K04-1}. An MPS is a kind of nested Schmidt decomposition and its time evolution
is computed without decomposing the data structure. Thus, an MPS simulation is fast for quantum circuits
that do not increase the Schmidt rank significantly. The time complexity of MPS simulation of an $n$-qubit
circuit is given by $O(ngr_{\rm max}^3)$ where $g$ is the number of single-qubit and two-qubit quantum gates
and $r_{\rm max}$ is the maximum Schmidt rank during the simulation.

It was shown in my previous articles \cite{AST-S13-1,AST-S13-2,AST-S14} that high-precision computing is
needed for avoiding numerical errors for simulating comparably large quantum circuits with the MPS method.
This is valid considering the machine epsilon as a function of the circuit width. Even a very simple quantum
circuit with $n-1$ control qubits and a single target qubit may produce a Schmidt coefficient $\sim 2^{(-n+1)/2}$
in the resultant state.\footnote{
For a typical example, consider the state $|+_{n-1}\cdots+_{1}\rangle|-_0\rangle$ with $|\pm\rangle=
(|0\rangle\pm|1\rangle)/\sqrt{2}$ and apply ${\rm C}^{n-1}-U$ to the state (here, ${\rm C}^{n-1}$
indicates a set of $n-1$ control qubits corresponding to a certain control sequence like
$0_{n-1}1_{n-2}1_{n-3}\cdots0_{2}1_{1}$ and $U$ is a certain single-qubit rotation). The resultant
state has Schmidt coefficients $\sqrt{1-\varepsilon}$ and $\sqrt{\varepsilon}$ with $\varepsilon
\sim 2^{-n+1}$. This kind of states appear after an oracle call in the Grover search and
related algorithms. Clearly, one cannot neglect a Schmidt vector with amplitude $\sqrt{\varepsilon}$.
} The machine epsilon in an MPS simulation should be exponentially small in the number
of qubits to maintain the accuracy of time evolution. (A similar discussion is found in Viamontes's
thesis \cite{Vthesis} on his QuIDDPro decision-diagram-based simulator that uses a decision diagram concepts
for simulating quantum computing.)

In contrast, there have been several results of other authors without employing high precisions, on the
numerical quantum circuit simulation using MPS and related data structure \cite{K04-1,K04-2,Wang17,Dang19,Liu21}.
They employed either the double precision \cite{K04-1,K04-2,Wang17,Dang19} or the mixed precision not
exceeding the double precision \cite{Liu21}.
Thus, there is an essential conceptual difference in the precision management among the authors.
To my understanding, one cannot easily guarantee the correctness of numerical results without using
multiple-precision computing. Furthermore, inaccurate time evolution may result in a rapid increase
in the number of Schmidt coefficients.

In this report, I will introduce recent improvements in the extensions of my libraries for parallel
processing with OpenMP and CUDA. Speedup in Hermitian matrix diagonalization will be evaluated and
a typical quantum algorithm will be numerically simulated for evaluation.
\section{Extensions for Parallel Processing}
There are two extensions of the ZKCM library, ZKCM\_OMP and ZKCM\_CUS, for parallel processing
using OpenMP and NVIDIA CUDA, respectively. They are presently included in the development version of
ZKCM and found in subdirectories of the package.
\paragraph{ZKCM\_OMP}
The ZKCM\_OMP library provides a C++ name space in which several linear algebra routines are
implemented on the basis of the OpenMP framework together with multiple-precision libraries (i.e.,
GNU MP and MPFR, which are also basic libraries for the ZKCM main library). The main routines
are those for Gaussian elimination, LU decomposition, Hermitian matrix diagonalization, and singular
value decomposition.
\paragraph{ZKCM\_CUS}
The ZKCM\_CUS library provides a C++ name space in which the routines for Hermitian matrix
diagonalization and singular value decomposition are implemented on the basis of the NVIDIA CUSOLVER
library. The Rayleigh quotient iteration is utilized internally in these routines to recover the
precision after using double-precision calculations of CUSOLVER routines.
\subsection{Evaluation of Speedup}\label{subsec-speed}
The speed of Hermitian matrix diagonalization is especially important in the present context because
its cost is the dominant factor in the computational cost of the MPS simulation.
I previously reported \cite{AST-S13-1,AST-S14} that as for the speed of Hermitian matrix diagonalization,
ZKCM is comparable to PARI \cite{PARI}. In particular, I showed in Table 2 of Ref.~\cite{AST-S14} that 
the speed of the diag\_H function of ZKCM ver.0.3.6 is comparable or better than that of the eigen
function of PARI ver.2.5.3 when diagonalizing $50\times 50$ or larger-size Hermitian matrices for 768-bit precision.
Here, I evaluate the speed again using recent versions of ZKCM and PARI.

\begin{figure}
\begin{center}
\includegraphics*[angle=270,width=0.7\textwidth]{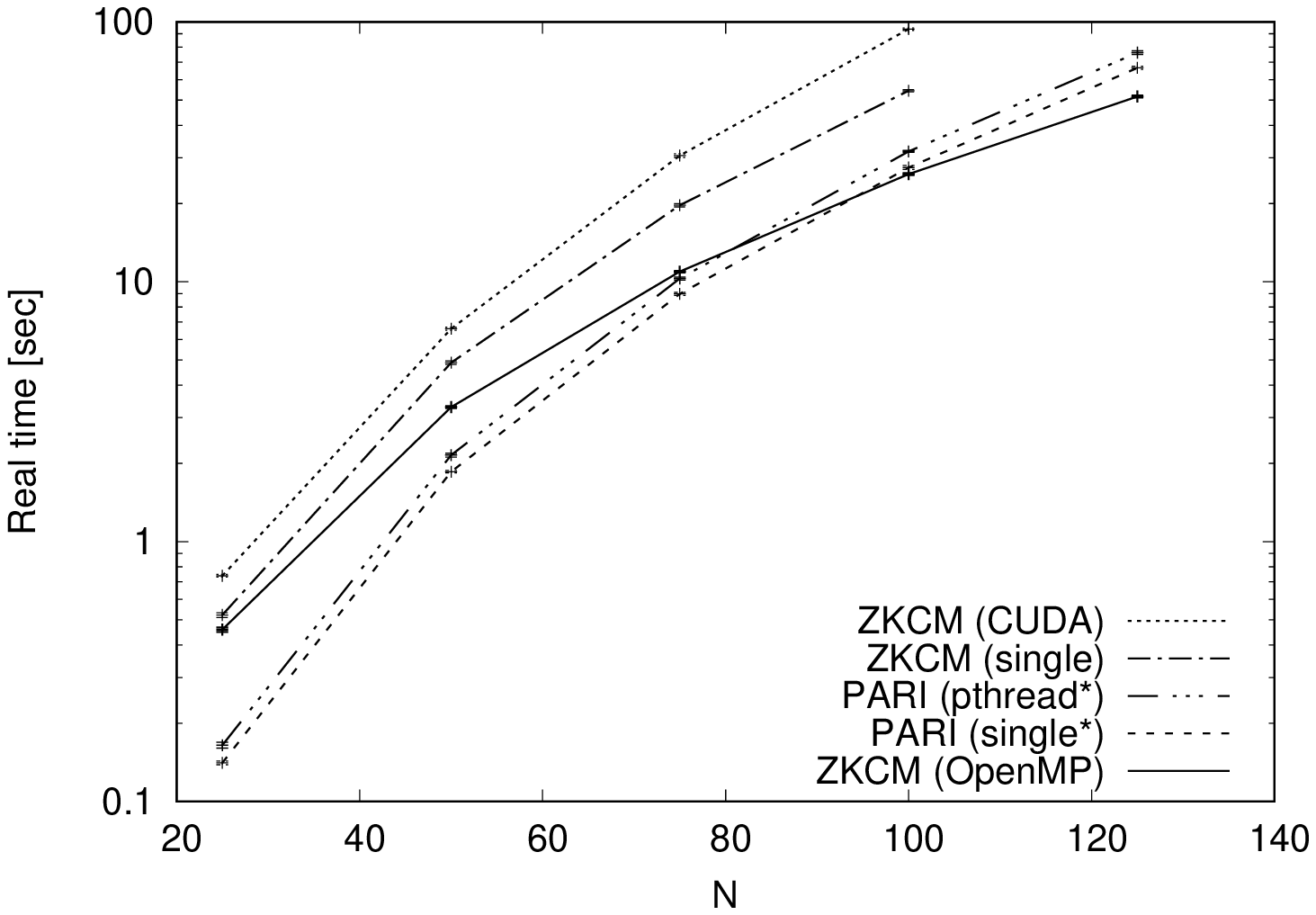}
\caption{\label{fig-diag}Comparison of computing time for Hermitian matrix diagonalization.
See the text for explanation.}
\end{center}
\end{figure}

Figure \ref{fig-diag} shows the results of computing time for diagonalizing $N\times N$ Hermitian matrix with
random elements. The 768-bit precision was used. Each data point is a 10-trial average and its error bar is the
standard deviation. The computation for "ZKCM (OpenMP)" involved 12 CPU threads. For PARI, it should be
noted that it does not have a routine specific to multi-thread matrix diagonalization and hence
the single/pthread switching did not make practical sense. The machine environment was as follows:
AMD Ryzen 9 5900X 12-core, $\sim$3.45GHz CPU, 64GB memory, NVIDIA GeForce GTX TITAN Black GPU.
The software environment was as follows: AlmaLinux 8.4 OS, GMP 6.1.2, MPFR 3.1.6, CUDA 11.7.1.
The versions of ZKCM and its extensions are as follows: ZKCM 0.4.4beta, ZKCM\_OMP 0.0.4alpha,
ZKCM\_CUS 0.0.9beta, and ZKCM\_QC 0.2.3beta.
The version of PARI was 2.15.4. In the figure,
"ZKCM (CUDA)" corresponds to the \verb|zkcm_cus::cus_diag_H| function,
"ZKCM (single)" corresponds to the \verb|diag_H| function,
"ZKCM (OpenMP)" corresponds to the \verb|zkcm_omp::diag_H| function, and
"PARI (pthread*)" and "PARI (single*)" correspond to the \verb|eigen| function; these
are function names for matrix diagonalization.

According to this result, the recent improvement in the diagonalization routine of PARI seems
rather significant and its performance exceeds that of ZKCM as far as quite recent versions are
used. For the matrix size $\ge 75$, however, the routine of ZKCM\_OMP showed faster speed
although we should also consider that it used 12 CPU threads.

From another point of view, the speedup by OpenMP is certainly observed for "ZKCM (OpenMP)"
in comparison to "ZKCM (single)".
The disappointing result was that "ZKCM (CUDA)" showed speed-down unexpectedly. This should
not be stressed much since this was owing to the machine environment; the CPU was in the
recent generation (year 2020) while the GPU was in the old generation (year 2014).

\subsection{Usage in an MPS simulation}
The recent development version of ZKCM\_QC has configuration options to use ZKCM\_OMP
and ZKCM\_CUS for Hermitian matrix diagonalization and singular value decomposition
when they are effective. A programmer does not have to pay attention to OpenMP or
CUSOLVER usage when properly-configured ZKCM\_QC was installed.
Sample codes are available in the samples directory of the ZKCM\_QC package (the package is linked
from URL \cite{AST-ZKCM}).
\section{Simulations of Quantum Algorithms}
\subsection{Deutsch-Jozsa Algorithm}
\begin{figure}
\begin{center}
\includegraphics*[angle=0,width=0.7\textwidth]{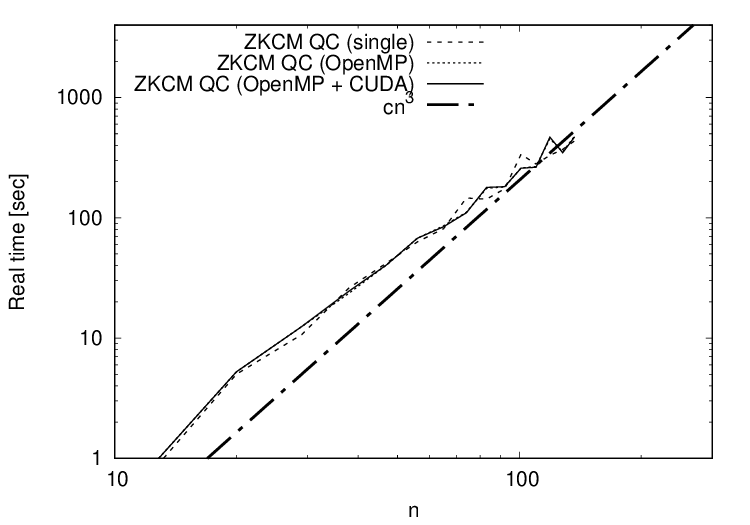}
\caption{\label{fig-DJ}Comparison of computing time for simulating quantum circuits of the
Deutsch-Jozsa algorithm for a certain function structure. $n$ stands for the total number of qubits
(namely, the circuit width).
The bold line indicates the line of ${\rm const} \times n^3$ with fitting to the data points
of "OpenMP+CUDA".
"single" stands for the case where ZKCM\_QC is compiled without parallel processing;
"OpenMP" stands for the case with ZKCM\_OMP;
"CUDA" stands for the case with ZKCM\_CUS;
"OpenMP+CUDA" stands for the case with both of ZKCM\_OMP and ZKCM\_CUS.
}
\end{center}
\end{figure}

The Deutsch-Jozsa algorithm is one of well-known quantum algorithms (see Ref.~\cite{NC2000}).
In my previous reports \cite{AST-S13-1,AST-S14}, a fast simulation of the algorithm was shown
for certain function structures using ZKCM\_QC. Here, the same function structure as Ref.~\cite{AST-S14}
was employed and simulation time was evaluated when OpenMP and CUDA CUSOLVER were present. 
The machine and software environment was identical to the one described in section \ref{subsec-speed}
except for that the 256-bit precision was employed and non-averaged raw data were used for the plots.
Figure \ref{fig-DJ} shows the time consumption results as functions of the circuit width $n$.
This suggests that no particular speedup by parallel processing was observed
although speedup by OpenMP was expected in light of the result shown in figure \ref{fig-diag}.
It was found that the matrix size handled by the diagonalization routines during the simulations was
at most $56\times 56$ and hence the speedup by parallel processing was very limited.
It is expected that the benefit of parallel processing will be clear when a more complicated function
structure causing a large Schmidt rank is involved.

\subsection{Remarks on Quantum Factoring}
It is controversial if a fast MPS simulation of Shor's quantum factoring circuit is possible.
In 2014, I reported \cite{AST-S14} a (practical-time\footnote{This indicates 24 hours or less, usually.})
simulation up to 54-qubit width using a single CPU thread with 128-bit precision in a PC workstation.
In 2017, Wang et al. \cite{Wang17} reported a simulation up to 45-qubit width using multiple
machine nodes with double precision (namely 53-bit precision) in a distributed computing framework.
In late 2017, I presented \cite{AST-17} a tentative result for 60-qubit width using a 128-bit precision
CPU thread partly with CUDA GPGPU computing in a PC workstation, which has been an unpublished work.
In 2019, Dang et al. \cite{Dang19} reported a simulation up to 60-qubit width under a similar
computing framework as Ref.~\cite{Wang17}.
To my understanding, high-precision computing is necessary to distinguish the space for non-zero
Schmidt coefficient from the null space to maintain the Schmidt rank as small as possible during
simulation. This, however, involves slow arithmetic computing for keeping the precision, and it is
not very clear if all non-zero Schmidt vectors contribute to the final result of the quantum circuit.
Therefore, a further investigation for larger factoring circuits is required to find which direction
is economical. A multiple-thread multiple-precision simulation is under progress and its result is
expected to appear in near future.
\section{Conclusion}
A recent progress in the ZKCM and ZKCM\_QC libraries have been described, in particular about the
extension libraries for parallel processing using OpenMP and CUDA CUSOLVER.
Clear speedup by OpenMP was observed in high-precision Hermitian matrix diagonalization.
It is expected that this will be beneficial for high-precision MPS simulation of quantum circuits
in case a relatively large Schmidt rank is involved although clear affirmative results could not
be shown in the present report.
\subsection*{Acknowledgement}
This work was supported by the Grant-in-Aid for Scientific Research from JSPS
(Grant No. 18K11344).

%
%

%
\end{document}